# Modulation-doped β-(Al$_{0.2}$Ga$_{0.8}$)$_2$O$_3$/ Ga$_2$O$_3$ Field-Effect Transistor

Sriram Krishnamoorthy [1, a, b], Zhanbo Xia [1, a, b], Chandan Joishi [1, 2], Yuewei Zhang [1], Joe McGlone [1], Jared Johnson [3], Mark Brenner [1], Aaron R. Arehart [1], Jinwoo Hwang [3], Saurabh Lodha [2] and Siddharth Rajan [1, 3, b]

[1] *Department of Electrical & Computer Engineering, The Ohio State University, Columbus, OH 43210, USA*
[2] *Department of Electrical Engineering, Indian Institute of Technology Bombay, Mumbai 400 076, India*
[3] *Department of Materials Science and Engineering, The Ohio State University, Columbus, OH 43210, USA*

[a] Sriram Krishnamoorthy and Zhanbo Xia contributed equally to this work

[b] Authors to whom correspondence should be addressed.
Electronic mail: krishnamoorthy.13@osu.edu, xia.104@osu.edu, rajan.21@osu.edu

## Abstract

Modulation-doped heterostructures are a key enabler for realizing high mobility and better scaling properties for high performance transistors. We report the realization of modulation-doped two-dimensional electron gas (2DEG) at β-(Al$_{0.2}$Ga$_{0.8}$)$_2$O$_3$/ Ga$_2$O$_3$ heterojunction using silicon delta doping. The formation of a 2DEG was confirmed using capacitance voltage measurements. A modulation-doped 2DEG channel was used to realize a modulation-doped field-effect transistor. The demonstration of modulation doping in the β-(Al$_{0.2}$Ga$_{0.8}$)$_2$O$_3$/ Ga$_2$O$_3$ material system could enable heterojunction devices for high performance electronics.

Beta- Gallium Oxide[1,2] (GO) is a promising ultra-wide band gap material with a band gap of 4.6 eV. Its large band gap and availability of bulk substrates grown using melt-based growth techniques[3,4,5,6,7,8,9,10,11,12] make it attractive for high power[13], high frequency, and optoelectronic[14,15] applications. Field effect transistors[16,17,18,19,20,21] and diodes with up to 1 kV[22] breakdown have been demonstrated by several groups using both beta and other polymorphs of GO. Most transistor results in this material system[16-21)] have been based on field-effect transistors with relatively thick channels, using Schottky or metal-oxide gate structures for charge modulation. However, as in conventional III-V semiconductors, two-dimensional electron gas (2DEG) channels provide several advantages for scaling of devices, as well as for higher mobility[23]. In particular, in the case of GO, where the room-temperature bulk mobility is relatively low, the screening of phonon scattering in high density 2D electron gases has been predicted to lead to significantly higher mobility[24]. This high mobility combined with the vertical scaling enabled by 2DEGs would be very beneficial for a range of high performance device applications that exploit the large band gap of GO. Recently, such a scaled thin channel approach using delta impurity doping was used to realize a delta-doped FET[25] with high transconductance and current density. A modulation-doped structure would enable higher mobility than a delta-doped layer by reducing the effect of impurity scattering.

$Al_2O_3$ is stable in the corundum polytype while $Ga_2O_3$ has multiple polytypes, with the monoclinic crystal structure (β-$Ga_2O_3$) being the most stable. Alloying of Al with β-$Ga_2O_3$ to realize β-$(Al_xGa_{1-x})_2O_3$ (AGO) has been investigated[26] and $Al_2O_3$ mole fraction as high as 61% was reported while maintaintaining phase purity for molecular beam epitaxy-grown films[27]. $Al_2O_3$ mole fraction up to 80% have been reported for AGO powders prepared using solution combustion synthesis[28] and pulsed laser deposition[29]. Evidence of modulation doping has also been reported in at the AGO/GO heterojunction using unintentional doping of the AGO[30]. In this work, we report on the electrical characteristics of a delta modulation-doped epitaxial AGO/GO heterostructure and demonstrate a modulation-doped AGO/GO modulation-doped field effect transistor (MODFET).

An AGO/GO modulation-doped heterostructure (epitaxial stack shown in Figure 1(a)) was grown on Fe-doped (010) β-Ga$_2$O$_3$ semi-insulating substrates[31]. The epitaxial structure consisted of 125 nm unintentionally-doped (UID) GO buffer layer and a 21 nm AGO barrier layer with delta doping in the AGO barrier. AGO spacer layer thickness of 3 nm was used to maximize the modulation doping efficiency. Before initiating the growth, the substrate was heated up to 800 °C to remove adsorbed impurities on the substrate to avoid any parasitic channel in the buffer layer. The sample was grown[32] by oxygen plasma- assisted molecular beam epitaxy using a Ga flux of 8 x 10$^{-8}$ Torr, Oxygen plasma power of 300W, and chamber pressure of 1.5 x 10$^{-5}$ Torr. AGO was grown[33] using an Al flux of 1.6 x 10$^{-8}$ Torr while keeping Ga flux at 8 x 10$^{-8}$ Torr. For the delta-doped layer, the Si shutter was opened for 7 seconds (0.25 nm), with the Si cell temperature set to 925º C. While multiple delta-doped layers were used to achieve high charge density in our earlier report[25], we used a single sheet of delta-doped layer with a lowered target doping density of ~ 9.5 x 10$^{12}$ cm$^{-2}$ in this work. The doping density was lowered to enable complete modulation of the 2DEG charge. High resolution X-ray diffraction (HRXRD) measurements was used to determine the composition of the AGO barrier layer (Fig. 1(b)). Assuming pseudomorphic growth, the peak separation between the AGO and the Ga$_2$O$_3$ peak was used to determine the Al composition[34] of the AGO barrier layer as 20%. The full width half maximum (FWHM) value of the AGO peak was measured to be 0.24 deg, which is higher than the typical thickness-related broadening observed in AlGaN/GaN heterostructures for the same barrier layer thickness (0.2 deg FWHM for AlGaN peak), indicating that the crystalline quality of the AGO layer can be improved further. Atomic force microscopy (Fig. 1(c)) indicates smooth surface morphology with an rms roughness of 0.7 nm. The cross-sectional scanning transmission electron microscope (STEM) image of a sample grown using identical growth conditions is shown in Fig. 1 (d). The STEM image clearly shows the presence of AGO/GO heterojunction with a nominal thickness of 20-21 nm for the AGO layer.

Ohmic contacts were formed using a side-metal contact, where the ohmic regions were etched down to 40 nm, thick Ti (60 nm)/ Au (50 nm)/ Ni (100 nm) metal was deposited, and annealed at 470°C in $N_2$ ambient for 1 min in a rapid thermal anneal system. Mesa isolation was carried out using $BCl_3$- based ICP-RIE etch, and gate metal stack of Ni/Au/Ni was deposited on the sample surface to form a Schottky barrier contact. Hall measurements were carried out using patterned van der Pauw structures, and sheet charge density of $5 \times 10^{12}$ cm$^{-2}$ and mobility of 74 cm$^2$/Vs were measured (sheet resistance of ~ 16.7 kΩ/square).

Capacitance-voltage characterization (Fig. 2(a)) (10 kHz) showed characteristic accumulation behavior indicating the presence of a 2D electron gas. A total sheet charge density of $8.5 \times 10^{12}$ cm$^{-2}$ was extracted from the CV curve. Anti-clockwise hysteresis of approximately 0.1 V was observed in double-sweep CV measurements suggesting motion of charged ions or defects within the structure. Further characterization is required to understand the observed hysteresis behavior. No frequency dispersion was observed between CV measurements at 1 kHz and 10 kHz. CV measurements could not be performed at higher frequencies due to the high series resistance in this structure.

The discrepancy between the charge density measured using Hall measurements and the CV charge is attributed to the incomplete ionization of donors in AGO at equilibrium ($V_{GS} = 0$

V). Delta doping in the AGO barrier layer results in reduction of seperation between AGO conduction band edge and the fermi level. This results in a reduced energy difference between the donor energy level and the fermi level, leading to incomplete ionization of donors at equilibirum. When a negative gate bias is applied, the conduction band in the delta-doped AGO layer and the donor energy level are pulled up relative to the fermi level, resulting in increased ionization of donors. The donor ionization process continues with increased negative bias until all the donors are completely ionized. When the donors are completely ionized, further increase in negative gate bias voltage depletes the 2DEG at the heterojunction. Hence CV characterization is expected to overstimate the 2DEG equilibrium charge density. Furthermore, modulation of the ionized donors with gate bias results in a parasitic capacitance parallel to the 2DEG capacitance affecting the extraction of 2DEG characterisitics (charge location) exclusively from CV measurements .

The apparent charge concentration profile extracted from CV characteristics is shown in Fig.2 (b). We have taken into account the spread of dielectric constant (10 [35,36]- 13 [37]) reported in the litreature. It should be noted that the dielectric constant of GO/AGO along the (010) orientation has not been characterized yet. The charge profile extracted from the CV measurements clearly shows charge confinement. Location of the apparent charge profile peak (assuming $\varepsilon_r = 13$) at the dopant site is due to the effect of the parasitic capacitance resulting from donor ionization, which is confirmed from CV simulation explained below.

The equilibrium energy band diagram (Fig. 3(a)) was calculated using self-consistent Schrodinger-Poisson simulations (bandeng)[38], for a nominal $Al_2O_3$ molefraction of 20% for the AGO barrier. Material parameters ($\varepsilon$ - relative permittivity, $m^*_e$ – effective electron mass, $\Phi_B$ – AGO surface barrier, $\Delta E_c$ – AGO/GO conduction band offset) were based on previous experimental and theoretical estimates and are listed in Table 1. The Schottky barrier height was extracted to be 1.4 eV using internal photoemission measurements [see Supplementary material Fig. S1]. The calculated equilibrium sheet charge (5 x $10^{12}$ cm$^{-2}$) matches well with the experiment (Hall measurement) if an AGO/GO conduction band offset of 0.6 eV and a donor energy level of 135 meV ($E_C - E_D$) is assumed in AGO. This would indicate that the band gap difference[27] appears completely as conduction band offset in AGO/GO heterojunction. The Si donor sheet density in the delta-doped layer was set to be 9.5 ×$10^{12}$ cm$^{-2}$. The simulated CV curve matched well with the experiments when the dielectric constant was assumed to be 13[37]. Further detailed experiments are required to measure band offsets and donor energy level in AGO.

TABLE I. Material parameters used for AGO/GO band diagram simulation

| Material Property | Value |
|---|---|
| Ga$_2$O$_3$ bandgap | 4.6 eV |
| (Al$_{0.2}$Ga$_{0.8}$)$_2$O$_3$ bandgap | 5.2 eV [29] |
| $\Delta E_c$ | 0.6 eV |
| $\varepsilon$ | 10 [35,36] – 13 [37] |
| $E_C - E_D$ | 135 meV |
| $m^*_e$ | 0.27 |
| $\Phi_B$ | 1.4 eV |

Electrical characterisitcs of the AGO/GO MODFET is shown in Fig. 4. Due to the non-ohmic source/drain contacts (Supplementary information Fig. S2) and high contact resistance, the transistor characterisitcs were measured on I-shaped structures consisting of wide source/drain contact/access regions ($W_{contact}$ =100 µm) and a narrow channel region ($W_{channel}$ = 2 µm) , (Fig. 4 (a)). The maximum current (Figure 4 (a)) normalized with respect to the channel width was measured to be 5.5 mA/mm. Measurements on devices with different channel width and length dimensions clearly indicate that the measured current density is rather limited by the contacts (Fig. S3, Table S1). The pinch-off voltage of -3 V measured in the transfer characteristics (Fig. 5(b)) was similar to the pinch-off voltage in the CV characteristics, and an on-off ratio of 2.5 x $10^5$ was obtained. A peak transconductance of 1.75 mS/mm was measured. While this report of modulation doping in such a transistor structure shows the feasibility of this approach, further growth optimization would be needed to achieve the optimal mobility.

In summary, we report modulation doped $(Al_{0.2}Ga_{0.8})_2O_3/Ga_2O_3$ heterostructures with evidence of carrier transfer from the $(Al_{0.2}Ga_{0.8})_2O_3$ layer to the $Ga_2O_3$ layer. Early results on field effect transistors confirm the presence of a 2D electron gas. Further work on improving the transport and resistance in such devices could enable the realization of high performance heterostructure devices based on the AGO/GO material system.

**Supplementary Material**:
See supplementary material for internal photoemission characterization of the $Ni/(Al_{0.2}Ga_{0.8})_2O_3$ barrier height, source/drain contact characterization, and the effect of device width, length on maximum drain current of AGO/GO MODFET.

**Acknowledgement**:
We acknowledge funding from Office of Naval Research, Grant Number N00014-12-1-0976 (EXEDE MURI). We acknowledge funding from The Ohio State University Institute of Materials Research (IMR) Multidisciplinary Team Building Grant. We thank Air Force Research Laboratory, WPAFB, Dayton Ohio for support.

**Figure Captions**

**Figure 1**: (Color online) (a) Epitaxial structure , (b) High resolution X-ray diffraction characterization, (c) Atomic force microscopy image of a typical sample surface, and (d) cross-section STEM image of AGO/GO MODFET grown on insulating Fe-doped GO substrate.

**Figure 2**: (Color online) (a) Experimental capacitance-voltage characteristics of AGO/GO MODFET, (b) apparent electron concentration profile extracted from CV measurements for dielectric constant values of 10 and 13.

**Figure 3**: (Color online) (a) Experimental capacitance-voltage characteristics of AGO/GO MODFET, (b) apparent electron concentration profile extracted from CV measurements for dielectric constant of 10 and 13.

**Figure 4**: (Color online) (a) Output characteristics and (b) transfer characteristics of AGO/GO MODFET sample C showing FET operation with charge modulation, pinch-off and a high ON/OFF ratio.

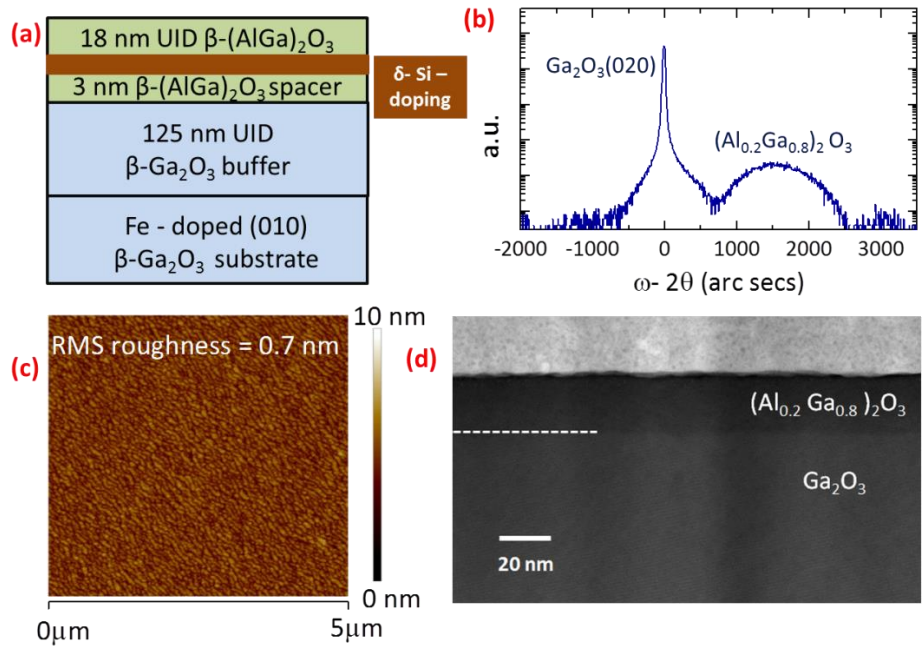

**Figure 1**

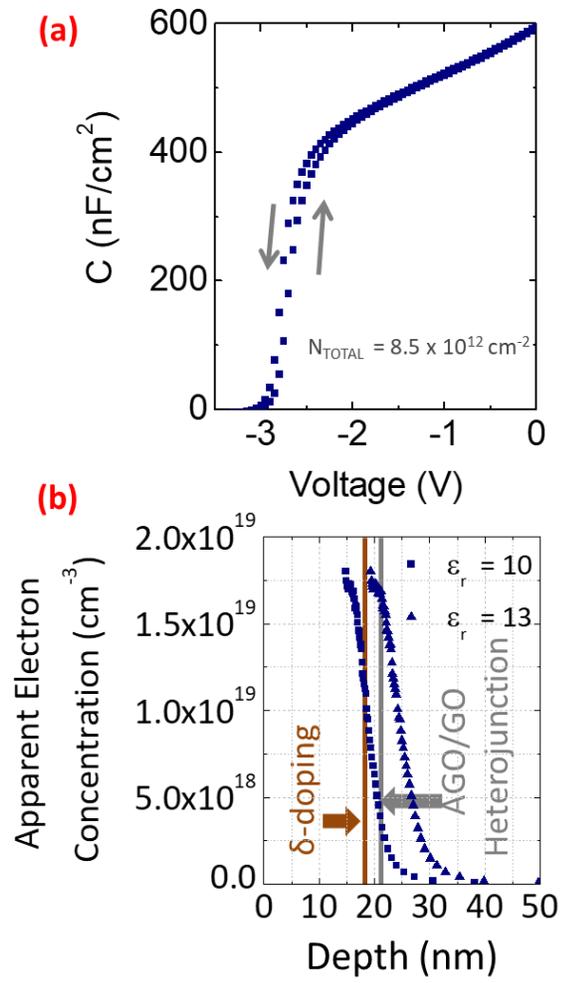

**Figure 2**

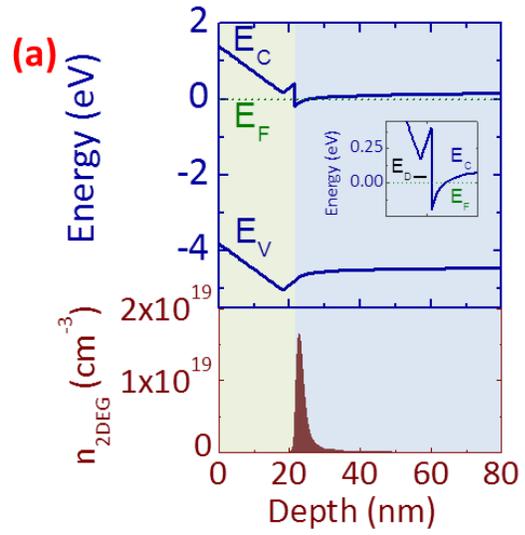

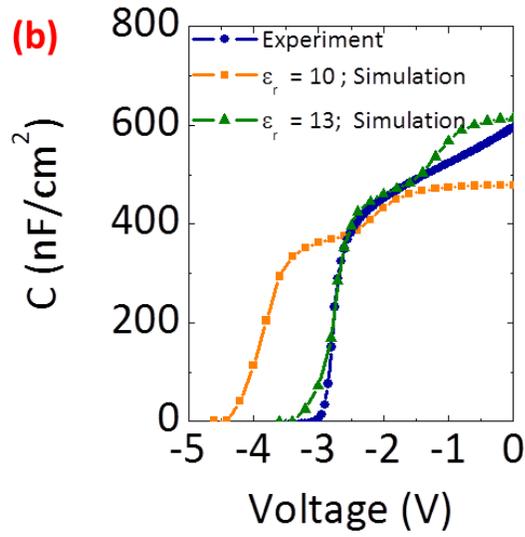

**Figure 3**

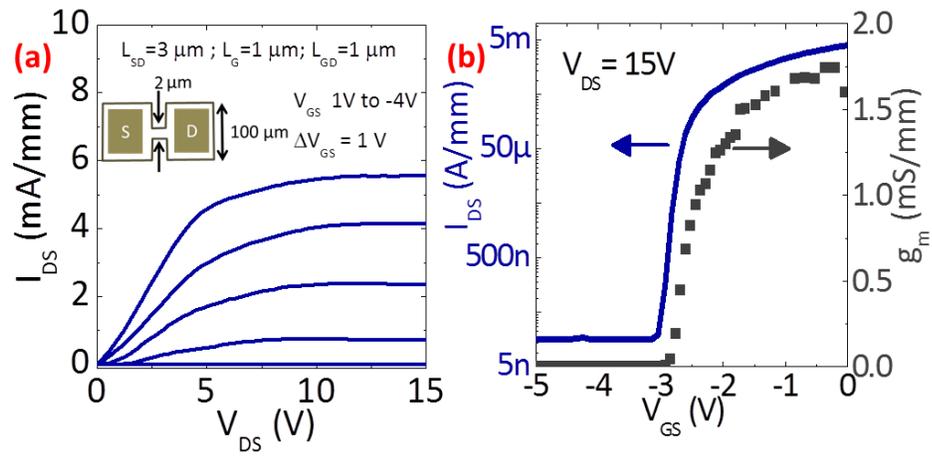

**Figure 4**